\documentclass[preprint,prd,showpacs,showkeys,twoside,aps]{revtex4}

\renewcommand{\Re}{\mathrm{Re\:}}
\renewcommand{\Im}{\mathrm{Im\:}}
\usepackage{amssymb}
\usepackage{graphicx}
\begin{document}
 \title{Representation of Integral Dispersion Relations by Local Forms }
\author{Erasmo Ferreira}
\affiliation{Instituto de F\'{\i}sica, Universidade Federal do Rio
de Janeiro, 21941-972, Rio de Janeiro , Brasil }
\author{Javier  Sesma}
\affiliation{Departamento de F\'{\i}sica Te\'orica, Facultad de
Ciencias, 50009, Zaragoza, Spain }
\begin{abstract}

 The representation of the usual integral dispersion relations
(IDR) of scattering theory through series of derivatives of the
amplitudes is discussed, extended, simplified, and confirmed as
 mathematical identities.
  Forms of derivative dispersion relations (DDR)
valid for the whole energy interval, recently obtained and
presented as double infinite series,
 are simplified through the use of new sum
 rules of the incomplete $\Gamma$ functions, being reduced to single
 summations, where the usual convergence
 criteria are easily applied.
 For the forms of the imaginary amplitude used in phenomenology of
 hadronic scattering at high energies, we show that expressions for
the DDR  can represent, with absolute accuracy, the IDR of
scattering theory, as true mathematical identities.
  Besides the fact that the algebraic manipulation can be easily
  understood, numerical examples show the accuracy of these
   representations up to the maximum available machine precision.

    As consequence of our work, it
    is concluded that the standard forms sDDR,
     originally intended for high energy limits,
    are an inconvenient and incomplete separation of terms of
    the full expression, leading to wrong evaluations.

  Since the correspondence between IDR and the DDR
  expansions is linear, our results have wide applicability,
  covering more general  functions, built as combinations of
  well studied basic forms.

\end{abstract}
\pacs{13.85.Dz, 13.85.Lg, 13.85.-t} \keywords{hadronic scattering;
dispersion relations; high energies}
\maketitle

\section{Introduction}

We here deal with the one-subtracted integral dispersion relations
(IDR) used in high energy scattering, as introduced originally   
 to study the properties of the complex
functions that represent the scattering amplitudes
\cite{idr1,idr2}. These relations are  derived from general
principles of causality, analyticity and crossing,
 and provide connections between the real and imaginary parts of
 the amplitudes, helping in the disentanglement of the
 expressions that represent the observable probabilities.
 The relations are written for amplitudes that have specific
 behavior under the operation of crossing symmetry, that connects
 $a+b \rightarrow a+b$ and $a+\bar b \rightarrow a+\bar b$
 reaction channels  ($\bar b$ is the antiparticle of $b$).

 These dispersion relations are derived using the Cauchy theorem,
 having the form of infinite integrations over the energy
 of the scattering process.  The applications in phenomenology
 are restricted because experimental information is limited
 to finite energy values. In spite of this limitation,
 applications of IDR  to the analysis of hadronic scattering
 were numerous in the years that followed the original work.

 The conversion of integrations over infinite
energy intervals into infinite series of derivatives of the
amplitudes in a single energy leads to the so called derivative
dispersion relations (DDR). The program has run through a long and
rather slow progress, which started in 1969  \cite{fddr}, when, in
a specific dynamical model in Regge theory framework, Gribov and
Migdal wrote a relation in which  real and imaginary amplitudes
are connected by a single derivative with respect to the logarithm
of the energy.

More general relations of local form were written for the even and
odd (under crossing) amplitudes \cite{bks}, through algebraic
transformations and series expansions that required appropriate
analytical properties of the imaginary amplitude. 
At about the same time, similar derivative dispersion relations
were written \cite{kn} and applied to the analysis of hadronic
scattering within the Regge dynamics approach. The expressions 
of these DDR are formed by infinite series of derivatives 
of increasing order, with the form of a symbolic tangent operator
in the derivatives with respect to the logarithm of the energy,
as written below in Eqs. (\ref{ddrplus_new}, \ref {ddrminus_new}).

The need of mathematical convergence of the series,
which depends on the analytic form of the imaginary amplitudes,
 has restricted  the applicability and credibility of the method, 
and soon it was analysed in both aspects of mathematical validity 
and physical usefulness. Examples of physical interest were given
 \cite{ed} for which the series are divergent, as in the presence
of resonance poles in general, and in particular for a Breit-Wigner
amplitude.  

Convergence conditions were discussed by Heidrich and Kazes 
\cite{hk} who proved that the representation of the amplitude 
as a series in the logarithm of the energy must have an infinite 
convergence radius; thus the amplitude must be an entire function 
of the energy logarithm. Again it was here observed that in the 
presence of singularities the DDR representation is not justified.

The use of DDR with a Regge representation of the amplitudes,   
consisting only of log, log-squared and power terms of the energy,
without resonance poles, first introduced by Kang and Nicolescu 
\cite{kn}, was discussed by Bujak and Dumbrajs \cite{bd}. The authors 
warn for the spurious singular behavior of the tangent series for
specific values of the power parameters, which is a peculiar feature
of the series summation in the original form of DDR obtained in the 
low energy approximation. 

Concerns with the mathematical validity of the DDR lead 
Fischer and Kol\'a\v{r}, in a series of papers \cite{kf,kola,kf-blois},  
to the establishment of conditions under which the series of 
derivatives have meaning, converging at a point or in an energy range. 
 
The DDR technique was considered of practical usefulness, in spite 
of the limitations  on the form of functions to which it could be
applied, and it was actually used in phenomenological
investigations in pp and p$\bar {\rm p}$ scattering. Limitations
in this use of DDR were due to the lack of knowledge of the
importance of the approximations involved in their derivation, of
which the so called high energy approximation is the most
important, and also the role of a free parameter $\alpha$ that
appeared in the algebraic manipulation leading from IDR to DDR
\cite{bks}. Use of the integration constant $K$ that appears in
the one-subtracted IDR, and sometimes of the value of the quantity
$\alpha$, as free parameters, filled gaps between phenomenological
parameterizations and data, hiding some of the difficulties.

A comprehensive account of the progress of DDR, with clear
description of limitations and successes of the method in its
applications to hadronic (pp and  p$\bar {\rm p}$)  scattering, is
found in \cite{am04}.
As far as the applications  to hadronic scattering are concerned,
practical uses of DDR  only consider  a few forms of energy
dependence of the imaginary amplitude that are required to
represent the data  \cite{compete}.

A main step in the progress came with the extension of the DDR
 to the low energy region, up to the threshold \cite{am07},
 with a  derivation free of the high energy
approximation, opening the way to the true mathematical connection
between integral and local representations of scattering
amplitudes. The expressions given for the DDR are double infinite
series, with rather complicated form, but really presenting no
practical difficulty for their numerical summation, as the series 
have quickly decreasing terms.
The only remaining difficulty may rest in the characterization of
convergence criteria for double series. Applications to the
description of pp and p$\bar {\rm p}$ data have been very
successful, showing perfect superposition of evaluations of 
the real amplitudes through IDR and the full exact DDR.

In the present paper we deal with the establishment of
mathematical relations through which local expressions represent
infinite dispersion integrals exactly. We are not concerned here
with phenomenological fittings or choice of parameters so that we 
drop the additive constant $K$ and put the parameter $\alpha=1$ . 
Using sum rules of the incomplete gamma functions which have been 
derived before \cite{abad}, we are
able to reduce the double series to single summations, where
convergence properties are of elementary knowledge. The
connections thus established between principal value integrals and
series summations are true mathematical identities, and even lead
to the establishment of some new formulae of mathematical
interest. Our results are explicitly presented for all functions 
used in hadronic scattering phenomenology
\cite{am04,compete,am07}. Since the connections between integral
and local dispersion relations are linear, we can study simple
forms separately, without concern on their superposition to
construct physical amplitudes. For imaginary amplitudes of the
form of functions of  the energy logarithms, built as sums of 
products of the  type  $[\ln(E/m)]^k \times (E/m)^\lambda$, it seems 
that no restriction or approximation  needs to be introduced to
represent by single series the integral dispersion relations of
scattering theory. The convergence criteria are the same as the
conditions for validity of the original integral forms.

The standard forms of DDR, called sDDR, viciously affected  by the
high energy approximation, often have  simple expressions in terms
of elementary functions, but are frequently inaccurate as
representations of the original IDR that they intend  to
substitute. They even create spurious singular behavior where the
principal value integrals are regular. Our investigation of the
asymptotic behavior of the full DDR shows also that the sDDR may
even be incomplete as a high energy limit approximation.

Our work leads to the conclusion that the derivation of \'Avila 
and Menon \cite{am07} is  correct, and shows that the
separation between standard sDDR and correction terms is
unnecessary and misleading, and from now on should  be considered
only for historical reference.

The exact mathematical correspondence between integral and local
dispersion relations opens new lines of investigation, not only in
hadronic scattering, but also in other areas of physics where the
 use of dispersion techniques is important.

 In the present work we take as given the description of the
 historical facts and  critical analysis about derivative dispersion
 relations \cite{am04},  and start from the recent results of \'Avila
 and Menon \cite{am07} that obtained expressions for DDR valid for
 the whole energy  interval above the physical threshold.

  The paper  develops the analytical treatment of  cases of
  imaginary amplitudes of forms  $(\ln (E/m))^k\times (E/m)^\lambda$,
  where $k$ is an integer.
First considering imaginary amplitudes $\Im F_+$ and $\Im F_-$
  of form $(E/m)^\lambda$, where series of derivatives become
  series of usual functions, we recall a mathematical relation
 \cite{abad} that
  expresses  in closed form the sums over the incomplete $\Gamma$
  functions that appear in the double series. The resulting
  single series are studied in detail. As illustration of the 
  analytical results, numerical comparison
  of maximum machine accuracy between the principal value
  integral and the DDR representation is successfully performed.

   Special discussion is made of the behavior of DDR in the
  cases $k$ = 0, 1, 2  in the proximity  of values of
   $\lambda$ ($\lambda \rightarrow 0$) where spurious
  singularities appear in the expressions of sDDR. Through 
  numerical tables we show how the correction terms of the full 
  DDR introduce singularities of opposite sign that exactly 
  compensate the sDDR disease,
  leading to results that regularize the full DDR and are
  confirmed by the original IDR. 

\section{General formulae}

    We are here concerned with the study of  problems in the strong
  interaction of fundamental particles, such as proton-proton
 (pp)  and proton-antiproton ( p$\bar {\rm p}$ )  collisions at
 high energies. All masses being equal, these are processes of
  the kind   $m+m \rightarrow m+m$.
 Protons have antiprotons   as their antiparticles, and the pp and
   p$\bar {\rm p}$   channels are
    connected by the operation of crossing symmetry, leading to
   the definition of combined  amplitudes, which are even
   (subscript +) or odd (subscript $-$) under such operation.

    When dispersion relations are constructed, care must be taken
    of the behavior  at large distances in the complex plane,
     in order to find the proper function for which the relations
     are valid. The factors introduced as needed for convergence
     at infinity appear in the form of subtractions, which
     introduce external  constants. In the case of high energy
     scattering here considered, one subtraction is needed, and
     the Integral Dispersion Relations (IDR) are written
     \cite{idr1,idr2}
    \begin{equation}
    \label{inteven}
\Re F_+(E,t) = K
+\frac{2E^2}{\pi}P\int_{m}^{+\infty}dE^\prime \, \frac{ \Im
F_+(E^\prime,t)}{E^\prime(E^{\prime 2}-E^2)}
     \end{equation}
      and
\begin{equation}
\label{intodd}  \Re F_-(E,t)=
\frac{2E}{\pi}P\int_{m}^{+\infty}dE^\prime \, \frac{ \Im
F_-(E^\prime,t)}{(E^{\prime 2}-E^2)} \, ,
\end{equation}
respectively for the amplitudes which are even and odd under
crossing. $K$ is the subtraction constant of the one-subtracted
dispersion relation, $m$ is the proton mass, $E$ is the total
energy  of the incident particle in the laboratory system, and
$t$ is the momentum tranfer in the elastic collision ($t$ will 
 be written as equal to zero from now on).
  The pp and p$\bar {\rm p}$ channels are governed by
the combinations
$$ F_{\rm pp}= F_+ + F_- \,,\hspace{2cm} F_{\rm p \bar p}
           =F_+ - F_-  \, .  $$

This is a well defined theoretical framework, but it has  the
  practical difficulty that it requires the knowledge of
  imaginary  amplitude in the whole energy interval, from threshold
  $m$ to infinity, for the calculation of the real part.
  Extrapolations to far away energies go beyond safe
  phenomenological or model-building grounds.
    This difficulty lead to the development of local
  relations, in which the real amplitude at a given energy
  is expressed in terms of derivatives of the imaginary part
  at the same energy, or in a nearby range.

  With the notation and kinematics used in Eqs. (\ref{inteven}) and
(\ref{intodd}), these  DDR introduced for the high energy
conditions are written
\begin{equation}
\label{ddrplus_new} \Re F_+(E,t)= K  + E
 \tan \bigg(\frac{\pi}{2} \frac{d}{d\ln{E}}\bigg )
    \Bigg[\frac{ \Im F_+(E,t)}{E} \Bigg]  \, ,
\end{equation}
 \begin{equation}
\label{ddrminus_new}
 \Re F_-(E,t) = \tan   \bigg( \frac{\pi}{2}
\frac{d}{d\ln {E}}\bigg) \Im F_-(E,t) ~ .
\end{equation}

    These forms are derived for high  energies ($ m $ goes to zero
   or $ E $ goes to infinity). They are the traditional DDR, often
   applied in  the description
   of pp and  and  p$\bar {\rm p}$ cross sections \cite{am04},
   called {\it standard derivative dispersion relations}, sDDR,
    by  \'Avila and Menon \cite{am07}, to distinguish them from 
    extended  forms eDDR that aim at covering also the low energy 
   region, and which are discussed in the present work.

    The constant $K$, which is important in phenomenology, would 
    be transferred without any action from
    Eq. (\ref{inteven}) to the  final exact formulae that we write,
    and will be dropped from   now on.

\section {Conditions for convergence}

 The introduction of DDR raised concerns about its mathematical
 validity , due to the infinite series of operator derivatives of
 increasing order, symbolically represented by the tangent operator.

     In practice, since  we do not know the exact analytical forms
   of the theoretical amplitudes, in the applications of DDR at high 
   energies one deals with  phenomenological representations, obtained from
   fitting or from models, in terms of simple functions of $E$  and
   $\ln E$, used in the description of pp and  p$\bar {\rm p}$
   at high energies. For these  cases, the questions of convergence
    and mathematical validity can  be treated  directly,
    and may even become trivial, with explicit sums of the resulting
    series. Actually, in asymptotic limits the whole tangent series
    can eventually be reduced to its first term.

     However, in principle dignified relations should be written and
  valid for the exact amplitudes, whatever be their analytical forms.
  Eichmann and Dronkers \cite{ed} have shown that the DDR are valid only
  for certain classes of entire functions of $\ln (E/m)$.
  In a series of papers, Kol\'a\v{r} and Fischer \cite{kf,kola,kf-blois}
 established theorems fixing the conditions for the validity of the
   representations  of amplitudes in terms of series of derivatives.
     The main results are expressed as follows.

{\bf Theorem :}    Let $ f:R^1 \rightarrow  R^1$. The series
$$    \tan \, \bigg[ \frac{\pi}{2}\frac{d}{dx}  \bigg]\, f(x) $$
converges at a point $x$ in $R^1$ if and only if the series
$$    \sum_{n=0}^{\infty}\frac {d^{2n+1}}{dx^{2n+1}}\, f(x) $$
is convergent.

Besides, another theorem implies that the convergence in an energy
interval requires that $f(x)$ be an entire function of complex $x$
and that the series of positive orders in derivatives
        $$    \sum_{n=0}^{\infty}\frac {d^{2n}}{dx^{2n}}\, f(x) $$
 also converges.

  These  important theorems allow the examination of the convergence of the
  tangent series that appear in the applications of DDR to high
  energy scattering, where simple functions like $x$, $x^2$,
  $e^{\lambda \,  x}$ , $x \, e^{\lambda \,  x } $, ...  appear.
  Here $x$ is written in the place of $\ln (E/m)$.

\section { Low Energies }

   The high energy approximations restrict the use of DDR  in
   regions of a few GeV where models and different descriptions of
   pp  and p$\bar {\rm p}$ scattering  are  tested and compared.
   \'Avila and Menon \cite{am07} presented a  derivation of  DDR
   avoiding the high energy approximations, and showed  results
   representing well the exact IDR, even close  to the threshold
   ($E \approx m$). The results, presented in the form of double
   infinite sums, have their conditions of convergence difficult
   to prove.  In this section we test these results in some cases
   of interest.

The results of \'Avila and Menon are presented in the following
form, called eDDR (extended Derivative Dispersion Relations). For
even parity they write
\begin{equation}
\Re F_{+}(E)=
E\tan\left(\frac{\pi}{2}\frac{{\mathrm{d}}}{{\mathrm{d}}\ln
E}\right)\frac{\Im F_+(E)}{E} +\Delta^+(E,m)   \, ,
\label{eq:eddre}
\end{equation}
where the correction term $\Delta^{+}$ is given by
\begin{eqnarray}
\label{eq:deltap}
 \Delta^{+}(E,m) &=&
-\frac{E}{\pi}\ln\left|\frac{m-E}{m+E}\right|\frac{\Im F_+(m)}{m}  \\
&+&\frac{2E}{\pi} \sum_{k=0}^{\infty}\sum_{p=0}^{\infty}
\frac{(-1)^{k+1}\Gamma(k+1,(2p+1)\ln(E/m))}{(2p+1)^{k+2}k!}
\frac{d^{k+1}}{d (\ln E)^{k+1}} \frac{\Im F_+(E)}{E} \, .
\nonumber
\end{eqnarray}
 For the odd relation they obtain
\begin{equation}
\label{eq:eddrodd}
 \Re F_{-}(E)= \tan\left(\frac{\pi}{2}
\frac{{\mathrm{d}}}{{\mathrm{d}}\ln E}\right)\Im F_-(E)
+\Delta^-(E,m)  \, , \label{eq:eddro}
\end{equation}
where
 \begin{eqnarray}
 \label{eq:deltam}
\Delta^-(E,m) &=&
-\frac{1}{\pi}\ln\left|\frac{m-E}{m+E}\right|\Im F_-(m)      \\
&+& \frac{2}{\pi} \sum_{k=0}^{\infty}\sum_{p=0}^{\infty}
\frac{(-1)^{k+1}\Gamma(k+1,(2p+1)\ln(E/m))}{(2p+1)^{k+2}k!}
\frac{d^{k+1}}{d (\ln E)^{k+1}} \Im F_-(E)  \, . \nonumber
\end{eqnarray}

Equations (\ref{eq:eddre}) and (\ref{eq:eddro}) are meant to be
valid, for appropriate functional forms of the imaginary
amplitude, for any energy above the physical threshold $E = m$. As
a previous verification, before making use of our sum rules for
the $\Gamma$ functions, we present in this section practical examples
comparing, in accurate numerical terms, the original IDR (supposed
to be exact) with the improved eDDR. We take as working examples
the forms $(E/m)^\lambda$  for $\Im F_+$ and  $\Im F_-$,
and choose a very low energy $E=2m$ where the corrections are
expected to be large.  We then evaluate the contribution of each
term to the real  amplitude, through IDR and through sDDR and
eDDR, to check for  differences.

 To exemplify the use of Eqs. (\ref{eq:eddre}) and (\ref{eq:eddro}) in
  calculations in local form of the integral dispersion relations
  in Eqs. (\ref{inteven}) and (\ref{intodd}), we evaluate the
  differences between the RHS of Eqs. (\ref{inteven}) and
  (\ref{ddrplus_new}) and between the RHS 
        of Eqs. (\ref{intodd}) and (\ref{ddrminus_new})
        and compare with the correction terms  $ \Delta^+(E,m)$
        and $ \Delta^-(E,m)$ given in Eqs. (\ref{eq:eddre}) and
        (\ref{eq:eddro}) respectively.  With
        these forms for the imaginary amplitudes the tangent series
  of derivatives in sDDR are explicitly summable. We have in the even case
 \begin{equation}
 \label{power_even}
 \Re F_+ = (E/m)\tan \left(\frac{\pi}{2}\frac{d}{d\ln E}\right)
  (E/m)^{\lambda-1}= (E/m)^\lambda
\tan\bigg[\frac{\pi}{2}(\lambda-1 )\bigg]
\end{equation}
and in the odd case
 \begin{equation}
 \label{power_odd}
 \Re F_- =  \tan\left(\frac{\pi}{2} \frac{d}{d\ln E}\right) (E/m)^\lambda=
  (E/m)^\lambda \tan\bigg[\frac{\pi}{2}\lambda\bigg] \,  .
\end{equation}
 Results for
$\lambda$ equal to 0.5 , 0.75 and 1.0 are given in Table
\ref{data} below. The Cauchy Principal Value integrations were
made with the Mathematica 6.0  program. The case $\lambda=1$ is
not applicable to the odd case, as the integral in IDR diverges,
and also the sDDR gives the infinite value ${\rm Re} \, F_-=E \tan
(\pi/2)$ . The agreement is absolutely perfect, showing that for this 
example the expressions for
$\Delta^+(E,m)$ and $\Delta^-(E,m)$ \cite{am07} are the exact and
final corrections to sDDR.
  \begin{center}
   \begin{table}
   \caption{ \label{data} The low energy correction terms, with
   ${\rm Im}\, F_+=(E/m)^\lambda$, ${\rm Im}F_- \, =(E/m)^\lambda$, $E=2m$.
   The standard DDR gives wrong values for the real amplitudes of
   Eqs. (\ref{inteven}), (\ref{intodd}). The correction terms of
   Eqs. (\ref{eq:eddre}) and (\ref{eq:eddro})
   fill the gaps with all precision of the machine. }
   \vspace{.5 cm}
   \begin{tabular}{|c|c|c|c|c|c|c|c|c|}
   \hline
                & Re $F_+ $    & Re $F_+ $  & Expected        & $\Delta^{+}(E,m)$ \\
   $\lambda $   &  (IDR)      & (sDDR)      & correction     &                 \\
     \hline
     0.5        &-0.0665721174 &-1.4142135624&+1.3476414450   &+1.3476414450\\
     0.75       &+0.2202873309 &-0.6966213995&+0.9169087304   &+0.9169087304\\
     1.0        &+0.6993983051 & 0.0         &+0.6993983051   &+0.6993983051 \\
 \hline
 \hline
                & Re $F_- $    & Re $F_- $   & Expected        & $\Delta^{-}(E,m)$  \\
   $\lambda$    &  (IDR)      & (sDDR)      & correction      &                   \\
     \hline
     0.5        &+1.6536021594 &+1.4142135624&+0.2393885970    &+0.2393885970 \\
     0.75       &+4.2675819488 &+4.0602070605&+0.2073748883    &+0.2073748883 \\
   \hline
    \end{tabular}
   \end{table}
   \end{center}

\begin{figure}[th]
  \includegraphics*[height=7.5cm]{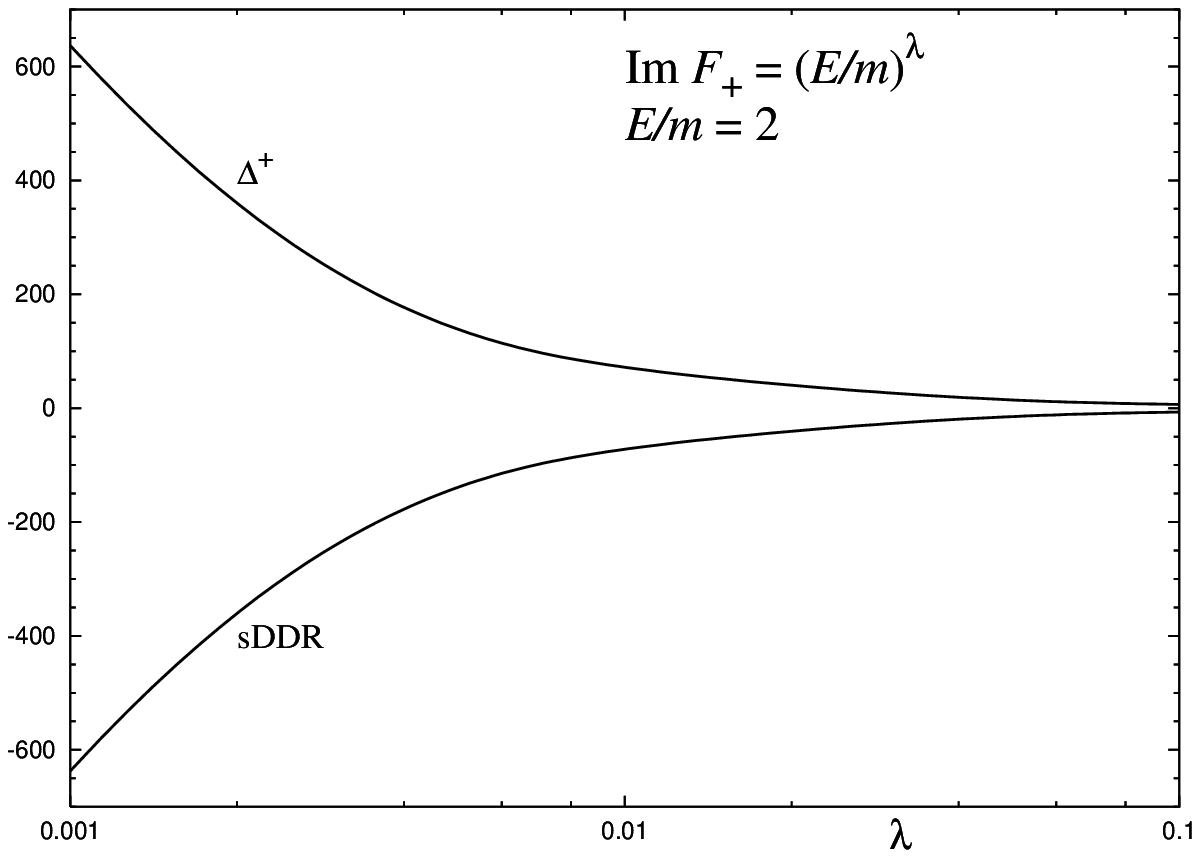} \hfill
 \caption{ sDDR and correction term $\Delta^+$ evaluated separately
 for the example $\Im F_+=(E/m)^\lambda$ in the vicinity of
 $\lambda=0$. The two parts become singular with opposite signs,
  while the sum (not shown in figure), giving the full DDR and the IDR
  is finite. The plot corresponds to the numbers given in Table
  \ref{tab2}, which shows that the full DDR gives precise description
  of the principal value integral. This example shows dramatically that
  the separation of sDDR and $\Delta$ terms is very inconvenient.}
\label{singular} \end{figure}

   The value $\lambda=0$ in $\Im F_+=(E/m)^\lambda$ is regular 
   in the IDR integral of Eq. (\ref{inteven}), but leads  to a
   singular value for sDDR of Eq. (\ref{power_even}). This
   spurious singularity in sDDR must be compensated by a
   similar singularity of opposite sign in $\Delta^+$. This is 
   actually the case, as exhibited numerically in Table \ref{tab2},
    and shown graphically in Fig. \ref{singular}.
\begin{center}
   \begin{table}
\caption{ \label{tab2} Real amplitudes from $F_+$ of  power form
   ${\rm Im}\, F_+= (E/m)^\lambda  $  at several energies, with
    $\lambda$ approaching the critical value  $\lambda=0$. The 
   standard DDR contribution and  correction terms $\Delta^+$ 
   added together reproduce perfectly the direct calculation  
   of the  IDR principal value   integral (calculated using the 
   Mathematica 6 software). For 
    $\lambda \rightarrow 0$ , sDDR and the corrections $\Delta^+$ have
   singularities of form $\approx 1/\lambda$ of opposite signs, that
   cancel in the sum. }
    \vspace{.5 cm}
   \begin{tabular}{|c|c|c|c|c|c|}
   \hline
   $\lambda $ &  $E/m$    &$\Re F_+$(IDR)&$\Re F_+$(sDDR)&$\Delta^{+}(E,m)$&$\Re F_+$(full DDR)\\
     \hline
 0.1  & 2 &-0.313205331434&-6.766911322477&6.453705991043& -0.313205331434 \\
  & 10 &-1.579297346713&-7.948542225578&6.369244878869& -1.579297346709\\
  & 100 &-3.640393754044&-10.006621794500&6.366228040456& -3.640393754044 \\
  & 1000 &-6.231392436474&-12.597590463309&6.366198026828&-6.231392436481\\
\hline
 0.01  & 2 &-0.346383461421&-64.099508812478&63.753125348801&-0.346383463676 \\
       & 10 &-1.474336661559&-65.139497143396&63.665160481840& -1.474336661556 \\
       & 100 &-2.994782036559&-66.656790947530&63.662008910971& -2.994782036558 \\
       & 1000&-4.547449511041&-68.209427064530&63.661977553485&-4.547449511046\\
\hline
 0.001  & 2 &-0.349370735776&-637.060672574268&636.711301837638&-0.349370736616\\
        & 10 &-1.463837806700&-638.086807702120&636.622969894093&-1.463837808012 \\
        & 100 &-2.937945523972&-639.557749708240&636.619804184254&-2.937945523972\\
        & 1000 &-4.412309900131&-641.032082585884&636.619772685732&-4.412309900136\\
 \hline
 0      & 2     &-0.349699152566&${\bf -\infty}$&${\bf +\infty}$&-0.349699152566\\
        & 10    &-1.462672076500&${\bf -\infty}$&${\bf +\infty}$&-1.462672076497\\
        & 100   &-2.931710562938&${\bf -\infty}$&${\bf +\infty}$&-2.931710562937\\
        & 1000  &-4.397613274962&${\bf -\infty}$&${\bf +\infty}$&-4.397613274967\\
 \hline
 \hline
    \end{tabular}
   \end{table}
   \end{center}
 The same behavior occurs in cases like $\ln (E/m) \,(E/m)^\lambda$
 and $(\ln (E/m))^2 \,(E/m)^\lambda$ which we have studied 
 similarly. Illustrative numerical tables are presented in the 
 appendix.

 In the next sections we show how to put in closed form one of
    the sums in $\Delta^+$ and $\Delta^-$, so that the convergence
    conditions can be studied  with the usual elementary methods.

\section {Reduction of the correction terms to single series: forms $(E/m)^\lambda$}

In terms of the dimensionless variable
\[
 \xi\equiv\ln(E/m)
\]
the \'Avila-Menon corrections \cite{am07} to the standard
derivative dispersion relations (sDDR) read
\begin{eqnarray}
\Delta^{+}(E,m) & = & -\frac{1}{\pi}\,\mbox{\rm e}^{\xi}\ln\left|\frac{1-\mbox{\rm e}^{\xi}}{1+\mbox{\rm e}^{\xi}}\right| \Im F_{+}(E=m)  \nonumber \\
 & & \hspace{-2cm}+ \frac{2}{\pi}\,\mbox{\rm e}^{\xi}\sum_{k=0}^{\infty}\sum_{p=0}^{\infty}\frac{(-1)^{k+1}\Gamma(k+1,(2p+1)\xi)}{(2p+1)^{k+2}\,k!}
\;\frac{d^{k+1}}{d\xi^{k+1}}\left(\mbox{\rm e}^{-\xi}\Im
F_{+}(E)\right)  \label{ap1}
\end{eqnarray}
and
\begin{eqnarray}
\Delta^{-}(E,m)& = & -\frac{1}{\pi}\,\ln\left|\frac{1-\mbox{\rm e}^{\xi}}{1+\mbox{\rm e}^{\xi}}\right| \Im F_{-}(E=m)  \nonumber\\
 & & \hspace{-2cm}+ \frac{2}{\pi}\,\sum_{k=0}^{\infty}\sum_{p=0}^{\infty}\frac{(-1)^{k+1}\Gamma(k+1,(2p+1)\xi)}{(2p+1)^{k+2}\,k!}
\;\frac{d^{k+1}}{d\xi^{k+1}}\left(\Im F_{-}(E)\right) \, .
\label{ap2}
\end{eqnarray}

The incomplete gamma function, with the first argument being an
integer, appearing in the right hand sides of both corrections, can
be written in the form \cite[p. 339]{magn}
\begin{equation}
\Gamma (k+1,z) = k!\, \mbox{\rm e}^{-z}\sum_{n=0}^k z^n/n!  \, ,
\label{ap3}
\end{equation}
from which it follows immediately the recurrence relation
\begin{equation}
\Gamma (k+1,z) = k\, \Gamma(k,z) + \mbox{\rm e}^{-z}\,z^k  ~ .
\label{ap4}
\end{equation}
This function admits the representations
\begin{equation}
\Gamma (k+1,z) =  \mbox{\rm e}^{-z}z^k \ _2\!F_0 (-k,1;;-1/z) ~ ,
\label{ap5}
\end{equation}
in terms of generalized Bessel polynomials, and
\begin{eqnarray}
\Gamma (k+1,z) & = & k! - \mbox{\rm e}^{-z}\,\frac{z^{k+1}}{k+1}\ _1\!F_1 (1;k+2;z) \label{ap6}\\
& = & k! - \frac{z^{k+1}}{k+1}\ _1\!F_1 (k+1;k+2;-z)\,,
\label{ap7}
\end{eqnarray}
in terms of confluent hypergeometric functions. These
representations may be useful, not only for the numerical
evaluation of the incomplete gamma function, but also for
algebraic simplifications of the summations in Eqs. (\ref{ap1})
and (\ref{ap2}), as we show below.

Important simplification of the double summations in the right
hand sides of Eqs. (\ref{ap1}) and (\ref{ap2}) may occur for
certain forms of $\Im F_+$ and $\Im F_-$. Let us consider the
expression
\begin{equation}
\mathcal{E}_p \equiv
\sum_{k=0}^{\infty}\frac{(-1)^{k+1}\Gamma(k+1,(2p+1)\xi)}{(2p+1)^{k+2}\,k!}
\;\frac{d^{k+1}}{d\xi^{k+1}}\left(\mbox{\rm
e}^{\lambda\xi}\right), \label{ap8}
\end{equation}
which appears when $F$ has a power form $E^\lambda$. Using the
representation (\ref{ap5}), it can be written in the form
\begin{equation}
\mathcal{E}_p = \frac{-\lambda \mbox{\rm
e}^{-(2p+1)\xi}}{(2p+1)^2}\,\mbox{\rm e}^{\lambda\xi}\,
\sum_{k=0}^{\infty}\frac{(-1)^{k}}{k!}\ _2\!F_0
\left(-k,1;;-1/(2p+1)\xi\right)\,(\lambda\xi)^k\,.
 \label{ap9}
\end{equation}
The sum rule
\begin{equation}
 \sum_{n=0}^{\infty} \frac{(-1)^n}{n!}\ _2\! F_0\!\left( -n,\gamma;;w\right) u^n
 = \mbox{\rm e}^{-u}\,(1-wu)^{-\gamma}\, , \qquad \mbox{\rm for} \quad |wu|<1 \, ,
 \label{ap10}
\end{equation}
a particular form of a known relation \cite[Sec. 19.10, Eq.
(25)]{erde}\cite[Eq. 6.8.1.2]{prud}
\begin{equation}
\sum_{k=0}^\infty\frac{t^k}{k!}\ _{p+1}\! F_q\!\left(
-n,(a_p);(b_q);x\right)= \mbox{e}^t\ _p\! F_q\!\left(
-n,(a_p);(b_q);-tx\right),  \label{cux10}
\end{equation}
that stems from a more general expansion of a hypergeometric
function in hypergeometric functions \cite[Eq. (1.6)]{fiel}
\begin{eqnarray}
\ _{p+r}\! F_{q+s} \left(
\left.\begin{array}{l}a_1,\ldots ,a_p,b_1, \ldots ,b_r \\
c_1,\ldots ,c_q,d_1, \ldots ,d_s\end{array}\right|zw\right)  =
\sum_{n=0}^{\infty}\frac{(a_1)_n\cdots(a_p)_n}{(c_1)_n\cdots(c_q)_n}
\,\frac{(-z)^n}{n!}  & & \nonumber \\  & &   \hspace{-9cm}
\ _p\! F_q\!\left(\left.\!\begin{array}{l}a_1\! +\! n,\ldots ,a_p\! +\! n \\
c_1\! +\! n,\ldots ,c_q\! +\! n\end{array}\right|z\!\right)
\ _{r+1}\! F_s\!\left(\left.\!\begin{array}{l}-n,b_1,\ldots ,b_r \\
d_1,\ldots \ldots ,d_s\end{array}\right|w\!\right), \label{cux11}
\end{eqnarray}
has been used  before \cite[Eq. (11)]{abad}, can be
checked by merely expanding its right hand side in powers of $u$,
and can be used to reduce to single summations the double sums
that appear in Eqs. (\ref{ap1}) and (\ref{ap2}). It allows us to
write
\begin{equation}
\mathcal{E}_p = -\lambda\,\mbox{\rm
e}^{\lambda\xi}\,\frac{\mbox{\rm e}^{-(2p+1)\xi}}{(2p+1)^2}
\left((1+\lambda/(2p+1)\right)^{-1}\,\mbox{\rm e}^{-\lambda
\xi}\,, \label{ap11}
\end{equation}
that is,
\begin{equation}
\mathcal{E}_p = -\lambda \,\frac{\mbox{\rm
e}^{-(2p+1)\xi}}{(2p+1)(2p+1+\lambda)}\, , \label{ap12}
\end{equation}
provided
\begin{equation}
\frac{|\lambda|}{2p+1} < 1\,. \label{ap13}
\end{equation}
Therefore, if the imaginary parts of the scattering amplitudes
have energy dependencies of the form
\begin{equation}
\Im F_{\pm} = A_{\pm}\,\exp (\beta_{\pm}\,\xi)\,, \label{ap14}
\end{equation}
the corrections given by Eqs. (\ref{ap1}) and (\ref{ap2}) can be
written respectively as
\begin{eqnarray}
\Delta^{+}(E,m)& = & A_+\Bigg[-\frac{1}{\pi}\,\mbox{\rm e}^{\xi}\ln\left|\frac{1-\mbox{\rm e}^{\xi}}{1+\mbox{\rm e}^{\xi}}\right| \nonumber \\
 & & \hspace{-2.5cm}+ \frac{2}{\pi}\,(1\!-\!\beta_+)\mbox{\rm e}^{(\beta_+-1)\xi}\sum_{p=0}^{p_+-1}\frac{\mbox{\rm e}^{-2p\xi}}{(2p+1)^2}\,
\sum_{k=0}^{\infty}\frac{(1-\beta_+)^k\xi^{k}}{k!}\ _2\!F_0 \left(-k,1;;\frac{-1}{(2p+1)\xi}\right)  \nonumber \\
 & & +\frac{2}{\pi}(1\!-\!\beta_+)\,\sum_{p=p_+}^{\infty}\frac{\mbox{\rm e}^{-2p\xi}}{(2p+1)(2p+\beta_+)}\Bigg] ~ , \label{ap15}
\end{eqnarray}
and
\begin{eqnarray}
\Delta^{-}(E,m) & = & A_-\Bigg[ -\frac{1}{\pi}\,\ln\left|\frac{1-\mbox{\rm e}^{\xi}}{1+\mbox{\rm e}^{\xi}}\right| \nonumber \\
 & & \hspace{-2cm}- \frac{2}{\pi}\,\beta_-\,\mbox{\rm e}^{(\beta_--1)\xi}\sum_{p=0}^{p_--1}\frac{ \mbox{\rm e}^{-2p\xi}}{(2p+1)^2}\,
\sum_{k=0}^{\infty}\frac{(-\beta_-)^k\xi^k}{k!}\ _2\!F_0 \left(-k,1;;\frac{-1}{(2p+1)\xi}\right)  \nonumber \\
 & & -\frac{2}{\pi}\beta_-\,\mbox{\rm e}^{-\xi}\sum_{p=p_-}^{\infty}\frac{\mbox{\rm e}^{-2p\xi}}{(2p+1)(2p+1+\beta_-)}\Bigg], \label{ap16}
\end{eqnarray}
where $p_+$ and $p_-$ represent the lowest integers verifying,
respectively,
\begin{equation}
2p_++1>|\beta_+-1|\,, \qquad 2p_-+1>|\beta_-|\,. \label{ap17}
\end{equation}
As a practical reference, we mention that a parametrization \cite{am07} of the
scattering amplitudes to fit the proton-proton and
antiproton-proton data is written
\begin{eqnarray}
\Im F_+(E) & = & X\,E^{\alpha_{P}(0)}+Y_+\,E^{\alpha_+(0)},  \label{ap18}\\
\Im F_-(E) & = & Y_-\,E^{\alpha_-(0)}.
\end{eqnarray}
Since the parameters giving the best fit verify the inequalities
\begin{equation}
1>|\alpha_{P}(0)-1|\,, \qquad 1>|\alpha_+(0)-1|\,, \qquad
1>|\alpha_-(0)|\,,  \label{ap19}
\end{equation}
Eq. (\ref{ap15}) with $p_+=0$ and Eq. (\ref{ap16}) with $p_-=0$
are applicable in this case. Then,
replacing $\ln |(1-\exp(\xi))/(1+\exp(\xi)|$ by its series
expansion, Eqs. (\ref{ap15}) and (\ref{ap16}) can be written
\begin{eqnarray}
\Delta^+(E,m) & = & A_+\,\frac{2}{\pi}\,\sum_{p=0}^{\infty}
\frac{\mbox{\rm e}^{-2p\xi}}{2p+\beta_+}\,, \qquad 0<\beta_+<2\,,\label{ap20} \\
\Delta^-(E,m) & = & A_-\,\frac{2}{\pi}\,\mbox{\rm
e}^{-\xi}\sum_{p=0}^{\infty}\frac{\mbox{\rm
e}^{-2p\xi}}{2p+1+\beta_-}\,, \qquad -1<\beta_-<1. \label{ap21}
\end{eqnarray}
Obviously, the series in the right hand sides of these equations
converge faster than the geometric one
$\sum_{p=0}^\infty(m^2/E^2)^p$ for any value of $\beta$, although
they contain singular terms when $\beta$ coincides with an even or
odd non positive integer. As it becomes clear from our derivation,
the restrictions on the values of $\beta$ in Eqs. (\ref{ap20}) and
(\ref{ap21}) are imposed to guarantee the applicability of the sum
rule Eq. (\ref{ap10}) in the replacement of the double sums in
(\ref{ap1}) and (\ref{ap2}) by the single sums in (\ref{ap20}) and
(\ref{ap21}).

\begin{figure}[th]
\includegraphics*[ height=7.5cm] {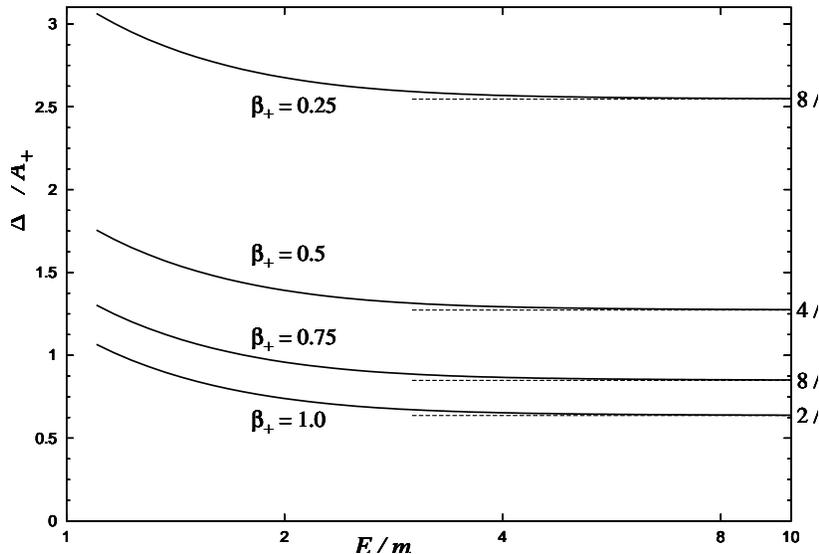} \hfill
  \caption{ Corrections to the even DDR, given by
   $ \Delta^+(E,m)$, for some values of $\beta_+$, as function of
      the energy. The nonzero values of $\Delta^+$ at high
      energies show that the sDDR do not give the correct and
      complete asymptotic forms. }
     \label{corr1}
      \end{figure}

\begin{figure}[th]
\includegraphics*[height=7.5cm] {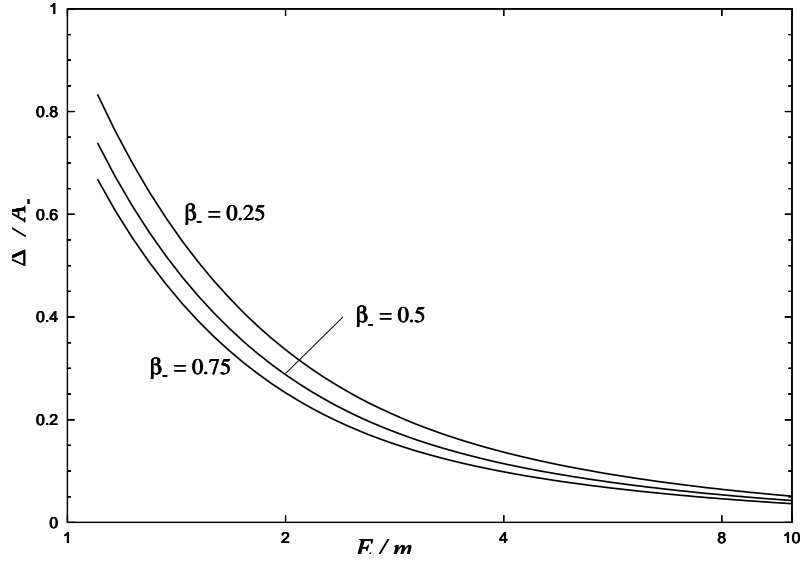} \hfill
  \caption{ Corrections to the odd DDR, given by
$\Delta^-(E,m)$, for some values of $\beta_-$, as function of
      the energy .    }
   \label{corr2}
      \end{figure}

   The values of $ \Delta^+(E,m)$  and  $ \Delta^-(E,m)$  for
   some values of $\beta$ are shown as functions of the energy in
   Figs. \ref{corr1} and \ref{corr2} respectively.
   The corrections for the even amplitude decrease with the energy to
   reach finite limits ${2}/{\pi\beta_+}$ asymptotically.
   The corrections $ \Delta^-(E,m)$ of the odd amplitude decrease
   to  zero as the energy increases.  The difference of behavior
    in the even and odd cases is mainly determined by the factor
    $e^{-\xi}=m/E$.
The behaviors of Eqs. (\ref{ap20}), (\ref{ap21}), in the limits of
large $\xi$ correspond to a  mathematical relation
\begin{equation}
\lim_{x_0 \to \infty}
 \Bigg[\,  \frac{2}{\pi}\, x_0^2 \,  P \int_{1}^{+\infty}
\!\frac{x^{\lambda-1} dx}{x^2-x_0^2}-x_0^\lambda \, \tan\left[
\frac{\pi}{2}(\lambda-1)\right] \Bigg]=\frac{2}{\pi\lambda} \, ,
\qquad
 0<\lambda <2 \; . \label{ap22}
\end{equation}

\begin{figure}[th]
\includegraphics*[ height=7.5cm] {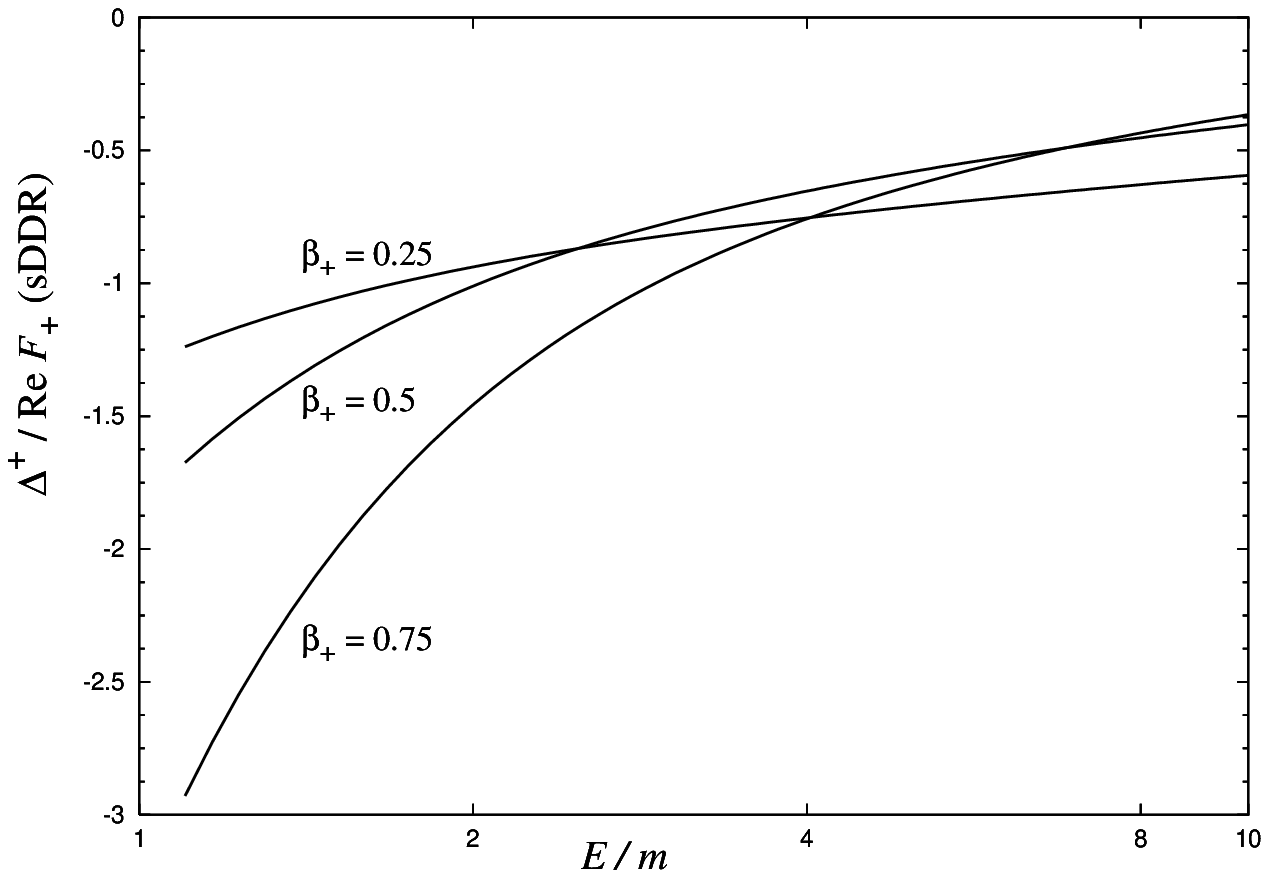} \hfill
\caption{  Ratios between the corrections to the even DDR, given by
   $ \Delta^+(E,m)$, and the standard high energy values given by
   sDDR , as functions of the energy, for a few values of $\beta_+$.}
    \label{corr3}
      \end{figure}

\begin{figure}[th]
\includegraphics*[ height=9.5cm] {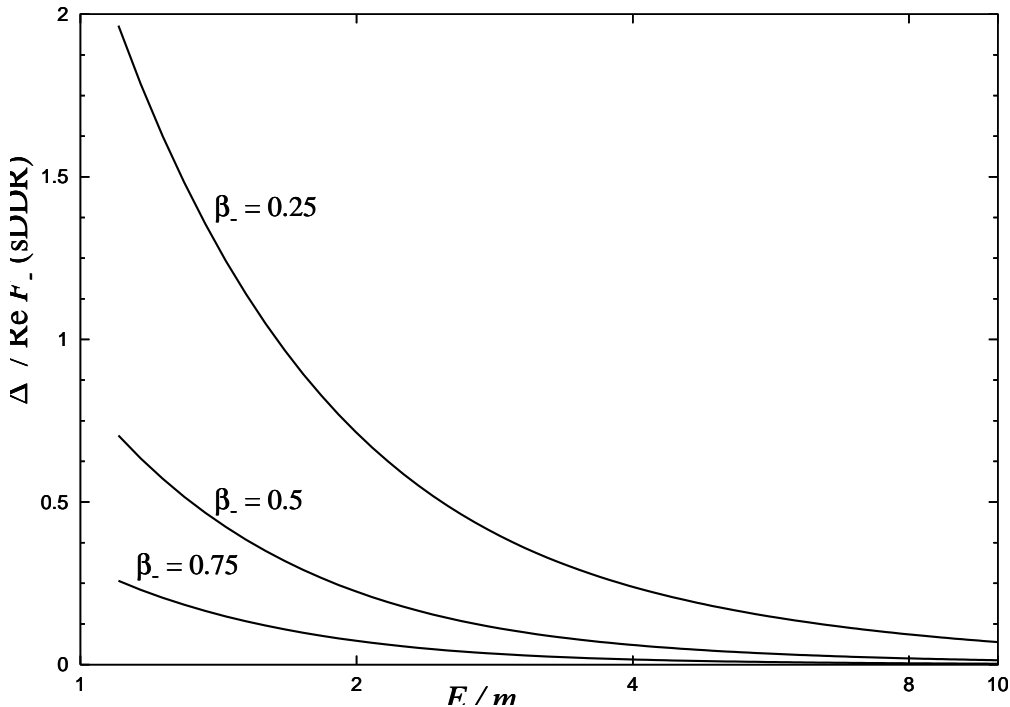} \hfill
\caption{ Ratios between the corrections to the odd DDR, given by
   $ \Delta^-(E,m)$, and the standard high energy values given by
sDDR , as functions of the energy, for a few values of $\beta_-$. }
      \label{corr4}
      \end{figure}

 To show the relative importance of the correction terms compared to the
 values of expressions given by sDDR, we draw in Figs. \ref{corr3} and
 \ref{corr4} the ratios  $\Delta^+(E,m)/{\rm Re}F_+({\rm sDDR)}$ and
      $ \Delta^-(E,m)/{\rm Re}F_-({\rm sDDR})$  as functions of the energy, for a few values of $\beta_+$ and $\beta_-$.

\section{ Full DDR for $\Im F$ written as combination of terms
$(E/m)^\lambda \,(\ln (E/m))^k$}

As mentioned before, sDDR suffer from serious deficiencies and
are not able to replace the IDR, even at  high energies. Addition
of the \'Avila-Menon corrections, to give what we call full
derivative dispersion relations, copes with those
deficiencies. We present, in this Section, a procedure to obtain
the full DDR in a general case of the imaginary part of the
amplitude being given by a sum
\begin{equation}
\Im F = \sum_{n} c_n\, (E/m)^{\lambda_n}\,(\ln (E/m))^{k_n},  \qquad
k_n \; \mbox{integer}, \label{dux1}
\end{equation}
with known coefficients $c_n$. If the sum is infinite, the series
must obey the convergence criteria mentioned before. 

The reason why we consider the form (\ref{dux1}) for the imaginary part of the
amplitude is mainly because it is an entire function of the logarithm of
the scattering energy, a condition that, as shown by Eichmann and Dronkers
\cite{ed}, guarantees the convergence of series like that appearing in
the sDDR, by expansion of the tangent function. Besides this, the finite
energy correction to the sDDR due to each term in the right hand side
of Eq. (\ref{dux1}) admits a very simple form: a single series, obviously 
convergent and easily calculable. We leave aside the crucial problem of the 
possibility of representing the scattering amplitude in the whole energy 
interval by sums of the form (\ref{dux1}). As pointed out also in 
Ref. \cite{ed}, singularities on the real $E$-axis due to inelastic 
thresholds and resonance poles for complex $E$ make difficult such 
representation with the sufficient accuracy as to avoid uncontrollable errors 
in $\Re F$, like those discussed in Ref. \cite[Sec. V]{kola}.
According to the comments of an anonymous Referee of a previous version of 
this paper, we do not solve the problem of finding a trustworthy representation 
of the scattering amplitude
as a superposition of entire functions of $\ln (E/m)$. The issue remains open.

The procedure to obtain full DDR in the case of $\Im F$ as given 
in Eq. (\ref{dux1}) is based on the linearity of dispersion relations 
and benefits from the fact that
\begin{equation}
(\ln (E/m))^n = \left. \frac{\partial^n}{\partial\lambda^n}
(E/m)^\lambda\right|_{\lambda=0}.  \label{dux2}
\end{equation}
We now focus our attention on the case of even scattering
amplitudes. The odd case is treated analogously.

We first consider the particular case
\begin{equation}
\Im F_+(E,m) = (E/m)^\lambda\,, \label{dux3}
\end{equation}
already discussed in previous Sections.
Equation (\ref {power_even}) for the even sDDR  gives
\begin{eqnarray}
\Re F_{+,\, sDDR}(E)  & = &
(E/m)^\lambda\,\tan\left[\frac{\pi}{2}(\lambda -1)\right]
  =  - \,(E/m)^\lambda\,\cot\left[\frac{\pi}{2}\lambda\right]   \nonumber \\
  & = &  (E/m)^\lambda\, \left[-\frac{2}{\pi\lambda}+\sum_{n=1}^\infty\frac{2^{2n}|B_{2n}|}
  {(2n)!}\,\left(\frac{\pi\lambda}{2}\right)^{2n-1}\right],  \nonumber \\
   & & \hspace{4cm} 0<\lambda<2, \label{dux5}
\end{eqnarray}
whereas the \'Avila-Menon correction, as obtained before, is
written
\begin{equation}
\Delta^+(E) = \frac{2}{\pi} \sum_{p=0}^{\infty}
\frac{(E/m)^{-2p}}{2p+\lambda}\,, \qquad 0<\lambda<2. \label{dux6}
\end{equation}
Adding the two contributions we obtain the full DDR for $\Im
F_+(E,m) = (E/m)^\lambda$,
\begin{eqnarray}
\Re F_+(E) & = & -\frac{2}{\pi\lambda}((E/m)^\lambda
-1)-(E/m)^\lambda\left[\cot\left(\frac{\pi\lambda}{2}\right)
-\frac{2}{\pi\lambda}\right]  \nonumber  \\ & & \hspace{1cm} \
+\frac{2}{\pi} \sum_{p=1}^{\infty} \frac{(E/m)^{-2p}}{2p+\lambda}\,,
\qquad -2<\lambda<2,\label{dux7}
\end{eqnarray}
or, equivalently,
\begin{eqnarray}
\Re F_+(E) & = & -\frac{2}{\pi}\sum_{n=0}^\infty\frac{\lambda^n(\ln
(E/m))^{n+1}}{(n+1)!}
+(E/m)^\lambda\sum_{n=1}^\infty\frac{2^{2n}|B_{2n}|}{(2n)!}\,\left(\frac{\pi\lambda}{2}\right)^{2n-1}
\nonumber \\  & & \hspace{1cm} \ +\frac{2}{\pi}
\sum_{p=1}^{\infty}\frac{(E/m)^{-2p}}{2p+\lambda}\,, \qquad -2<\lambda<2. \label{dux8}
\end{eqnarray}
Notice that the range of values of $\lambda$ for which the last
two equations are valid is larger than that of Eqs. (\ref{dux5})
and (\ref{dux6}). This is not surprising in view of the exact
cancellation, when summing, of the singularities occurring in the
right hand sides of Eqs. (\ref{dux5}) and (\ref{dux6}) for
$\lambda=0$. This fact has already been discussed at the end of
Section IV and is illustrated in Tables \ref{tab2} to \ref{tab4}.
Derivation ($k$-fold) with respect to $\lambda$ gives the full DDR
for $\Im F_+(E,m) = (E/m)^\lambda\,$ $(\ln (E/m))^k$ in the form,
\begin{eqnarray}
\Re F_+(E) & = & -\frac{2}{\pi}\, (\ln (E/m))^{k+1}\sum_{n=0}^\infty\frac{\lambda^n\,(\ln (E/m))^{n}}{n!\,(n+k+1)} \nonumber \\
 & & \hspace{-1cm}\ +\, \frac{2}{\pi}\,(E/m)^\lambda\sum_{j=0}^k \left(\!\begin{array}{c} k \\ j \end{array}\!\right) (\ln (E/m))^{k-j}\!\!\!
 \sum_{n=[j/2]+1}^\infty\!\!\!\frac{|B_{2n}|\pi^{2n}}{(2n)!}\,(2n\!-\!j)_j\,
 \lambda^{2n-1-j} \nonumber  \\
 & & \ +\, \frac{2}{\pi}\,(-1)^k\,k!\, \sum_{p=1}^{\infty}\frac{(E/m)^{-2p}}{(2p+\lambda)^{k+1}}\,,
 \qquad -2<\lambda<2. \label{dux9}
 \end{eqnarray}
 Alternatively, but much more tediously, the last equation could be obtained from Eq. (\ref{eq:eddre}) by direct computation and
 making use, to simplify the double series in Eq. (\ref{eq:deltap}), of sum rules stemming from Eq. (\ref{ap10})
 by successive derivations of both sides with respect to $u$.

 Taking $\lambda=0$ in (\ref{dux9}) we obtain obviously the
 full DDR for $\Im F_+(E,m) = (\ln (E/m))^k$,
 \begin{eqnarray}
\Re F_+(E) & = & -\frac{2}{\pi(k+1)}\, (\ln (E/m))^{k+1}  \nonumber  \\
 & & \ +\, \frac{2}{\pi}\,k!\,\sum_{l=1}^{[(k+1)/2]} \frac{(\ln (E/m))^{k+1-2l}}{(k+1-2l)!}
 \;\frac{|B_{2l}|\pi^{2l}}{(2l)!} \nonumber  \\
 & & \ +\, \frac{2}{\pi}\,(-1)^k\,k!\, \sum_{p=1}^{\infty}\frac{(E/m)^{-2p}}{(2p)^{k+1}}\,. \label{dux10}
 \end{eqnarray}
 It is now trivial to obtain the full DDR for $\Im F_+(E,m)$ as in Eq. (\ref{dux1}).

 In the case of odd amplitudes, proceeding as above, we obtain, for $\Im F_-(E,m) = (E/m)^\lambda \,(\ln (E/m))^k$,
 the full DDR expression
 \begin{eqnarray}
\Re F_-(E) & = & \frac{2}{\pi}\,(E/m)^\lambda\sum_{j=0}^k
\left(\!\begin{array}{c} k \\ j \end{array}\!\right) \! (\ln
(E/m))^{k-j}\!\!\!\!
 \sum_{n=[j/2]+1}^\infty\!\!\!\!\frac{(2^{2n}\!-\!1)|B_{2n}|\pi^{2n}}{(2n)!}\,(2n\!-\!j)_j\,
 \lambda^{2n-1-j} \nonumber  \\
 & & \ +\, \frac{2}{\pi} \,(-1)^k\,k!\, \sum_{p=0}^{\infty}\frac{(E/m)^{-2p-1}}{(2p+1+\lambda)^{k+1}}\,,
 \qquad -1<\lambda<1,
 \label{dux11}
 \end{eqnarray}
 and, taking in this expression $\lambda=0$, the full DDR for $\Im F_-(E,m) = (\ln (E/m))^k$
  \begin{eqnarray}
\Re F_-(E) & = &  \frac{2}{\pi}\,k!\,\sum_{l=1}^{[(k+1)/2]} \frac{(\ln
(E/m))^{k+1-2l}}{(k+1-2l)!}
 \;\frac{(2^{2l}-1)|B_{2l}|\pi^{2l}}{(2l)!} \nonumber  \\
 & & \ +\, \frac{2}{\pi} \,(-1)^k\,k!\,\sum_{p=0}^{\infty}\frac{(E/m)^{-2p-1}}{(2p+1)^{k+1}}\,. \label{dux12}
 \end{eqnarray}
 The exact DDR for $\Im F_-(E,m)$ of the form given by
 Eq. (\ref{dux1})  follows immediately.

For convenience of the reader, we present here the full DDR for the forms most frequently
 assumed for the imaginary part of the amplitude in the phenomenology of elementary particles,
 namely
 \begin{eqnarray}
 \Im F^{(a)} (E) & = & \ln (E/m)\,,  \\
 \Im F^{(b)} (E) & = & (\ln (E/m))^2\,,   \\
 \Im F^{(c)} (E) & = & (E/m)^\lambda\,,   \\
 \Im F^{(d)} (E) & = & (E/m)^\lambda\,\ln (E/m)\,,   \\
 \Im F^{(e)} (E) & = & (E/m)^\lambda\,(\ln (E/m))^2\,.
 \end{eqnarray}
 Full DDR for cases (a) and (b) can be obtained from Eqs. (\ref{dux10}) and (\ref{dux12}).
 The case (c) has been thoroughly discussed in the preceding Sections. Formulae for cases
 (d) and (e) stem from Eqs. (\ref{dux9}) and (\ref{dux11}). The resulting expressions are
 \begin{eqnarray}
 \Re F_+^{(a)} (E) & = & -\,\frac{1}{\pi}\,(\ln (E/m))^2+\frac{\pi}{6}-\frac{2}{\pi}\,
 \sum_{p=1}^\infty\frac{(E/m)^{-2p}}{(2p)^2}\,,   \\
 \Re F_-^{(a)} (E) & = & \frac{\pi}{2}-\frac{2}{\pi}\,
 \sum_{p=0}^\infty\frac{(E/m)^{-2p-1}}{(2p+1)^2}\,,  \\
 \Re F_+^{(b)} (E) & = & -\,\frac{2}{3\pi}\,(\ln (E/m))^3+\frac{\pi}{3}\,\ln (E/m)+\frac{4}{\pi}\,
 \sum_{p=1}^\infty\frac{(E/m)^{-2p}}{(2p)^3}\,,  \\
 \Re F_-^{(b)} (E) & = & \pi\,\ln (E/m)+\frac{4}{\pi}\,
 \sum_{p=0}^\infty\frac{(E/m)^{-2p-1}}{(2p+1)^3}\,  \\
 \Re F_+^{(c)} (E) & = & (E/m)^\lambda\,\tan\left[\frac{\pi}{2}(\lambda -1)\right]+
 \frac{2}{\pi} \sum_{p=0}^{\infty}\frac{(E/m)^{-2p}}{2p+\lambda}\,, \qquad |\lambda-1|<1\,, \label{fcp}\\
 \Re F_-^{(c)} (E) & = & (E/m)^\lambda\,\tan\left[\frac{\pi}{2}\lambda\right]+
 \frac{2}{\pi} \sum_{p=0}^{\infty}\frac{(E/m)^{-2p-1}}{2p+1+\lambda}\,, \qquad \qquad|\lambda|<1\,,   \\
 \Re F_+^{(d)}(E) & = & -\,\frac{2}{\pi}\,(\ln (E/m))^2\sum_{n=0}^\infty\frac{\lambda^n\,(\ln (E/m))^{n}}{n!\,(n+2)} \nonumber \\
 & & \ +\, \frac{2}{\pi}\,(E/m)^\lambda\,\sum_{n=1}^\infty\left[\lambda\,\ln
 (E/m)+2n-1\right]
 \,\frac{|B_{2n}|\pi^{2n}}{(2n)!}\,\lambda^{2n-2} \nonumber  \\
 & & \ -\, \frac{2}{\pi}\,\sum_{p=1}^{\infty}\frac{(E/m)^{-2p}}{(2p+\lambda)^2}\,, \hspace{4cm} |\lambda-1|<1\,, \\
 \Re F_-^{(d)}(E) & = & \frac{2}{\pi}\,(E/m)^\lambda\,\sum_{n=1}^\infty\left[\lambda\,\ln
 (E/m)+2n-1\right]
 \,\frac{(2^{2n}-1)|B_{2n}|\pi^{2n}}{(2n)!}\,\lambda^{2n-2} \nonumber  \\
 & & \ -\, \frac{2}{\pi}\,\sum_{p=0}^{\infty}\frac{(E/m)^{-2p-1}}{(2p+1+\lambda)^2}\,,
 \hspace{4cm} |\lambda|<1\,, \\
 \Re F_+^{(e)} (E) & = & -\,\frac{2}{\pi}\, (\ln (E/m))^3\sum_{n=0}^\infty\frac{\lambda^n\,(\ln (E/m))^{n}}{n!\,(n+3)} \nonumber  \\
 & & \ +\, \frac{2}{\pi}\,(E/m)^\lambda
 \sum_{n=1}^\infty\Big[\left[(\lambda\ln (E/m))^2+2(2n\!-\!1)\lambda\ln(E/m)+(2n\!-\!1)(2n\!-\!2)\right]  \nonumber \\
 & & \hspace{4cm} \frac{|B_{2n}|\pi^{2n}}{(2n)!}\,
 \lambda^{2n-3}\Big] \nonumber  \\
 & & \ +\, \frac{4}{\pi}\, \sum_{p=1}^{\infty}\frac{(E/m)^{-2p}}{(2p+\lambda)^{3}}\,,  \hspace{5cm} |\lambda-1|<1\,, \\
\Re F_-^{(e)} (E) & = & \frac{2}{\pi}\,(E/m)^\lambda
 \sum_{n=1}^\infty\Big[\big[(\lambda\ln (E/m))^2+2(2n\!-\!1)\lambda\ln(E/m)  \nonumber  \\
 & & \hspace{4cm} +(2n\!-\!1)(2n\!-\!2)\big]\frac{(2^{2n}-1)|B_{2n}|\pi^{2n}}{(2n)!}\,\lambda^{2n-3}\Big] \nonumber  \\
 & & \ +\, \frac{4}{\pi}\, \sum_{p=0}^{\infty}\frac{(E/m)^{-2p-1}}{(2p+1+\lambda)^{3}}\,~ ,  \hspace{3cm} |\lambda|<1\, . 
 \end{eqnarray}
 Notice that the right hand side of Eq. (\ref{fcp}) should be replaced by that of Eq. (\ref{dux8})
 for values of $\lambda$ in the interval $(-2,0]$.

\section{Contributions to the DDR for arbitrary forms of $\Im F$}

We now present alternative forms to the quantities $\Delta^+$ and
$\Delta^-$ of Eqs. (\ref{ap1}) and (\ref{ap2}), for arbitrary
functional forms of the scattering amplitudes, reduced through the
use of the sum rules of the incomplete gamma functions.
We first consider the even amplitude. The treatment of
corrections to the odd case is  similar.

Let us define the operator
\begin{equation}
\mathcal{V}(p) \equiv
\sum_{k=0}^{\infty}\frac{(-1)^{k+1}\Gamma(k+1,(2p+1)\xi)}{(2p+1)^{k+2}\,k!}
\;\frac{d^{k}}{d\xi^{k}} \, , \label{cux3}
\end{equation}
that, in view of Eq. (\ref{ap5}), can be written
\begin{equation}
\mathcal{V}(p) = -\, \frac{\mbox{\rm e}^{-(2p+1)\xi}}{(2p+1)^2}
\sum_{k=0}^{\infty}\frac{(-1)^{k}}{k!}\,\ _2\!F_0
\left(-k,1;;-1/(2p+1)\xi\right) \;\xi^k\,\frac{d^{k}}{d\xi^{k}} ~ .
\label{cux5}
\end{equation}
Since we are now dealing with derivative operators, the sum rule
in Eq. (\ref{ap10}) must be used carefully. For clarity, it should preferably
be written as
\begin{equation}
 \sum_{n=0}^{\infty} \frac{(-1)^n}{n!}\ _2\! F_0\!\left( -n,\gamma;;w\right) u^n
 = \mbox{\rm e}^{-u}\ _1\! F_0\!\left(\gamma;;wu\right)\,.  \label{cux9}
\end{equation}
To abbreviate, let us
define  the operators
 \begin{equation}
 D_{\xi} \equiv \frac{d}{d\xi}~ , \qquad \mathfrak{D}_\xi \equiv \xi\,\frac{d}{d\xi} \label{cux7}
 \end{equation}
 and their powers
  \begin{equation}
 D_{\xi}^n \equiv \frac{d^n}{d\xi^n} ~ , \qquad \mathfrak{D}_\xi^n \equiv \xi^n\,\frac{d^n}{d\xi^n}~ . \label{cux8}
 \end{equation}
 Notice that the last definition is an unnatural one. Then, from Eq. (\ref{cux5}) written,
 with this notation, in the form
\begin{equation}
\mathcal{V}(p) = - \,\frac{\mbox{\rm e}^{-(2p+1)\xi}}{(2p+1)^2}
\sum_{k=0}^{\infty}\frac{(-1)^{k}}{k!}\,\ _2\!F_0
\left(-k,1;;-1/(2p+1)\xi\right) \;\mathfrak{D}_\xi^k
\,,\label{cux12}
\end{equation}
one obtains, by using Eq. (\ref{cux9}),
\begin{equation}
\mathcal{V}(p) = - \,\frac{\mbox{\rm e}^{-(2p+1)\xi}}{(2p+1)^2}\,
\exp(-\mathfrak{D}_\xi)\ _1\!
F_0\!\left(1;;-\frac{D_\xi}{2p+1}\right) \,,\label{cux13}
\end{equation}
that, with the abbreviations
\begin{eqnarray}
\mathfrak{W}_1 & \equiv & \exp(-\mathfrak{D}_\xi) \nonumber \\
               & = & \sum_{n=0}^\infty \frac{(-1)^n\,\mathfrak{D}_\xi^n}{n!}, \label{cux14} \\
\mathfrak{W}_2(p) & \equiv & (2p+1+D_\xi)^{-1} \nonumber\\
               & = & \frac{1}{2p+1}\ _1\! F_0\!\left(1;;-\frac{D_\xi}{2p+1}\right) \nonumber \\
               & = & \frac{1}{2p+1}\sum_{n=0}^\infty \frac{(-1)^n}{(2p+1)^n}\,D_\xi^n ~ , \label{cux15}
\end{eqnarray}
can be written as
\begin{equation}
\mathcal{V}(p) = - \,\frac{\mbox{\rm
e}^{-(2p+1)\xi}}{2p+1}\;\mathfrak{W}_1 \,\mathfrak{W}_2(p)\, .
\label{cux16}
\end{equation}
Equations
(\ref{cux3}) and (\ref{cux16}) allow to write for the correction
to the sDDR for the even amplitude
\begin{eqnarray}
\Delta^{+}(E,m) & = & -\,\frac{1}{\pi}\,\mbox{\rm e}^{\xi}\ln\left|\frac{1-\mbox{\rm e}^{\xi}}{1+\mbox{\rm e}^{\xi}}\right| \Im F_{+}(E=m)  \nonumber \\
 & & - \,\frac{2}{\pi}\,\mbox{\rm e}^{\xi}\sum_{p=0}^{\infty}\frac{\mbox{\rm e}^{-(2p+1)\xi}}{2p+1}\;
 \mathfrak{W}_1 \,\mathfrak{W}_2(p)\,\frac{d}{d\xi}\left(\mbox{\rm e}^{-\xi}\,\Im F_+(E)\right),  \label{cux17}
\end{eqnarray}
that is, replacing the logarithm by its series expansion,
\begin{equation}
\Delta^{+}(E,m) = \frac{2}{\pi}\sum_{p=0}^{\infty}\frac{\mbox{\rm
e}^{-2p\xi}}{2p+1}
 \left[ \Im F_+(E=m)-\mathfrak{W}_1 \,\mathfrak{W}_2(p)\,\frac{d}{d\xi}\left(\mbox{\rm e}^{-\xi}\,\Im F_+(E)\right)\right],  \label{cux18}
\end{equation}
or, equivalently,
\begin{eqnarray}
\hspace{-1.5cm}\Delta^{+}(E,m) & = &
\frac{2}{\pi}\sum_{p=0}^{\infty}\frac{(E/m)^{-2p}}{2p+1}\;
 \Big[ \Im F_+(E=m) \nonumber \\
 & & \hspace{2cm}-\,\mathfrak{W}_1 \,\mathfrak{W}_2(p)\,E\,\frac{d}{dE}\left((E/m)^{-1}\,\Im F_+(E)\right)\Big] \, .  \label{cux19}
\end{eqnarray}
Following an analogous procedure, it can be obtained from Eq.
(\ref{ap2})
\begin{eqnarray}
\hspace{-2cm}\Delta^{-}(E,m) & = &
\frac{2}{\pi}\sum_{p=0}^{\infty}\frac{(E/m)^{-2p-1}}{2p+1}\;
 \Big[ \Im F_-(E=m)  \nonumber  \\
& & \hspace{2.5cm}-\,\mathfrak{W}_1
\,\mathfrak{W}_2(p)\,E\,\frac{d}{dE}\left(\Im F_-(E)\right)\Big].
\label{cux20}
\end{eqnarray}

An expression similar to Eq. (\ref{cux19}) has been given by Cudell, Martynov and
Selyugin \cite{cude,mart} to take account of the corrections to
the sDDR. Unfortunately, they do not give details of the derivation of their final
formula, that, besides containing an error, turns out to be confusing.
They write, for the corrections given in Eq. (\ref{cux19}), the expression
\begin{equation}
-\,\frac{2}{\pi}\sum_{p=0}^\infty \frac{C_+(p)}{2p+1}\left(\frac{m}{E}\right)^{2p}, \label{cud1}
\end{equation}
where
\begin{equation}
C_+(p)=\frac{\mbox{e}^{\xi D_\xi}}{2p+1+D_\xi}\left[\Im
F_+(E)-E\frac{d}{dE}\Im F_+(E)\right]. \label{cud2}
\end{equation}
Firstly, the absence in Eq. (\ref{cud1}) of a term like the first
one inside the bracket in the right hand side of Eq. (\ref{cux19})
suggests that those authors take for granted that $\Im F_+(E\! =\!
m)=0$. This is true for amplitudes of the form $\Im
F=(\ln(E/m))^k(E/m)^\lambda$, with $k$ positive integer, but not
for amplitudes like $(E/m)^\lambda$. Secondly, to avoid
ambiguities, the factor $(m/E)^{2p}$ in Eq. (\ref{cud1}) should be
moved to the left of $C_+(p)$, as this operator implies
derivations with respect to $E$. Thirdly, and more important, in
the expression (\ref{cud2}) of $C_+(p)$ one can find our operators
$\mathfrak{W}_1$ and $\mathfrak{W}_2$, defined in Eqs.
(\ref{cux14}) and (\ref{cux15}), but the order in which they act
on the functions at their right is not specified. Of course, the
order is not irrelevant, as it can be easily checked in a very
simple particular case:
\begin{eqnarray}
\mathfrak{W}_1 \,\mathfrak{W}_2(p)\,\mbox{e}^{\lambda\xi} & = &
\mathfrak{W}_1 \,\frac{1}{2p+1+\lambda}\,\mbox{e}^{\lambda\xi}
= \frac{1}{2p+1+\lambda}\,  \\
\mathfrak{W}_2(p) \,\mathfrak{W}_1\,\mbox{e}^{\lambda\xi} & = &
\mathfrak{W}_2(p) \,\mbox{e}^{0} = \frac{1}{2p+1}.
\end{eqnarray}
Finally, a factor $E^{-1}$ is lacking in front of the bracket in
the right hand side of Eq. (\ref{cud2}).

The usefulness of Eqs. (\ref{cux19}) and (\ref{cux20}) is
conditioned by the convergence of the series involved. As far as
we limit ourselves to forms of the type (\ref{dux1}), such
convergence is guaranteed, provided the parameters $\lambda_n$
satisfy the restrictions mentioned in the above equations. The
problem of the convergence for more general forms of $\Im F(E)$,
is beyond the scope of this paper. Nevertheless, the similitude of
the terms constituting the sDDR and those of the low energy
corrections, evident in the paper by \'Avila and Menon \cite[Sec.
II, Subsec. D]{am07}, together with Theorem 4 in Ref. \cite{kola},
allow us to conjecture that the converge conditions of the series
in $\Delta^+$ and $\Delta^-$ are the same as that of the tangent
series in sDDR, mentioned in our Section III.

It is interesting to write the expression of the sDDR in terms of
the operators introduced above. By expanding the tangent in a
double sum, one obtains
\begin{equation}
\Re F_{+,\,sDDR} = \frac{4}{\pi}\,(E/m)\,\sum_{p=0}^\infty
\sum_{k=0}^\infty\frac{1}{(2p+1)^{2k+2}}\,
\frac{d^{2k+1}}{d\xi^{2k+1}}\left(\mbox{e}^{-\xi}\,\Im F_+\right),
\label{cux21}
\end{equation}
that can be written in the form
\begin{equation}
\Re F_{+,\,sDDR} = \frac{4}{\pi}\,(E/m)\,\sum_{p=0}^\infty
\mathfrak{W}_2(p)\,\mathfrak{W}_3(p)
\,E\,\frac{d}{dE}\left((E/m)^{-1}\,\Im F_+(E)\right) \, ,
\label{cux22}
\end{equation}
with the operator $\mathfrak{W}_3(p)$ given by
\begin{eqnarray}
\mathfrak{W}_3(p) & \equiv & (2p+1-D_\xi)^{-1} \nonumber\\
               & = & \frac{1}{2p+1}\ _1\! F_0\!\left(1;;\frac{D_\xi}{2p+1}\right) \nonumber \\
               & = & \frac{1}{2p+1}\sum_{n=0}^\infty \frac{1}{(2p+1)^n}\,D_\xi^n. \label{cux23}
\end{eqnarray}
Analogously, one obtains for the odd case
\begin{equation}
\Re F_{-,\,sDDR} = \frac{4}{\pi}\,\sum_{p=0}^\infty
\mathfrak{W}_2(p)\,\mathfrak{W}_3(p) \,E\,\frac{d}{dE}\left(\Im
F_-(E)\right) \, . \label{cux24}
\end{equation}
Of course, the product of operators
$\mathfrak{W}_2(p)\,\mathfrak{W}_3(p)$ in Eqs. (\ref{cux22}) and
(\ref{cux24}) can be replaced by the operator
\begin{eqnarray}
\mathfrak{W}_4(p) & \equiv & ((2p+1)^2-D_\xi^2)^{-1} \nonumber\\
               & = & \frac{1}{(2p+1)^2}\ _1\! F_0\!\left(1;;\frac{D_\xi^2}{(2p+1)^2}\right) \nonumber \\
               & = & \frac{1}{(2p+1)^2}\sum_{n=0}^\infty \frac{1}{(2p+1)^{2n}}\,D_\xi^{2n}. \label{cux25}
\end{eqnarray}

\section{Conclusions}

 Local forms of dispersion relations have potential
 to become precious tools in theoretical physics. Avoiding
 the need of the knowledge of amplitudes and data at
 energies higher than experiments can provide, which
 limit the use of the integral dispersion relations,
 they become a useful technique in the interpretation
 of the experimental information  and in the
 formulation and test of theoretical models.

   Since its  origins \cite{bks}, the procedure to
pass from integral to local dispersion relations is based on
mathematical manipulations using series expansions and integrations,
whose validity are restricted by severe convergence criteria
\cite{ed,hk,kf,kola,kf-blois}. These requirements limit the forms of
the imaginary amplitude for which the procedure is mathematically
allowed. This limitation remains imperative, but it is not an
objection in the practice of hadronic phenomenology, since the forms
of amplitudes tested in the description of data are simple and
treatable functions of $E$ and $\ln E$. However, the predictions
obtained with the so called standard DDR, based on the high energy
approximation, can be absurdly wrong, 
as it was already pointed out and discussed by Bujak and Dumbrajs 
\cite{bd} and by Kol\'a\v{r} and Fischer \cite{kola}

Recently, \'Avila and Menon \cite{am07} reformulated the
transformation from integral to derivative dispersion relations
avoiding the high energy approximation. Using some examples, simple
and frequent in applications, we show in the present paper that
their new extended forms are complete representations.
The double infinite summations in the expressions are correct,
although apparently cumbersome, and in this form it is not trivial
to prove their convergence.

In the present work we went a step further in the development of
these extended forms of DDR, reducing the double to single series
summations for functions of general form  $ \Im F= (E/m)^\lambda
\times (\ln (E/m))^k$, with $k$ an integer, which may satisfy the
mathematical conditions in the passage from integral to local
dispersion relations. The resulting single summations are easy to
analyse in terms of usual convergence criteria, and easy to apply,
as the terms decrease rapidly.

We give explicit results for a number of basic forms and indicate
the way to more general possibilities. Through analytical and
numerical examples we show the practicability of the use of the 
exact local forms for the principal value integrals of IDR.

Our work shows that the so called standard derivative dispersion
relation, sDDR, cannot be trusted, and should not be used in
applications before specific tests, with comparison to the exact
DDR values. It is convenient, from now on, to work only with the
exact forms.

Since the connections from IDR to DDR are linear, results can be
combined linearly, and the treatment of the basic cases is
expected to be useful in many cases.

Our basic concern was with hadronic phenomenology, but we may hope
that the exact forms of DDR here discussed can be applied as a tool
and serve of inspiration in several other areas of physics. From
this development, we  believe that a new era of applications of
derivative dispersion relations can start.

 \begin{acknowledgments}
Financial support from Conselho Nacional de Desenvolvimento
Cient\'{\i}fico e Tecnol\'{o}gico  (CNPq, Brazil) and Comisi\'{o}n
Inter\-mi\-nis\-te\-rial de Ciencia y Tecnolog\'{\i}a (Spain) is
gratefully acknowledged. We are grateful to J. Abad, M. J.
Menon and R. F. \'Avila  for help and discussions. One of the 
authors (E.F.) is
very grateful to the Departamento de F\'{\i}sica Te\'orica of the
Facultad de Ciencias, Universidad de Zaragoza, for the friendly
hospitality during his stay in Zaragoza, when this work was made.
  \end{acknowledgments}

\appendix  \section{Numerical illustrations of use of DDR compared to IDR}

We present  in Tables \ref{tab3} and \ref{tab4} numerical  values 
of even IDR and DDR calculations of Re $ F_+$  , for imaginary amplitudes 
of forms  ${\rm Im}\, F_+= (E/m)^\lambda \ln(E/m) $
and ${\rm Im}\, F_+= (E/m)^\lambda \ln^2(E/m) $ ,  
 in the neighborhood of $\lambda=0$, where the spurious singularities
that appear in the separate parts sDDR and $\Delta^+$  are shown 
to cancel out, leading to the regular behavior of the full DDR, with 
accurate matching of the numerical principal value integrals. 
    \begin{center}
   \begin{table}
\caption{ \label{tab3} Real amplitudes from $F_+$ of  form
 ${\rm Im}\, F_+= (E/m)^\lambda \ln(E/m) $
 at several energies, with
    $\lambda$ approaching the critical value  $\lambda=0$. The standard 
    DDR contribution and  correction terms $\Delta^+$ added together 
    reproduce perfectly the direct calculation  of the  IDR principal value 
    integral (calculated using the Mathematica 6  software program). For 
    $\lambda \rightarrow 0$, sDDR and the corrections $\Delta^+$ have
   singularities of form $\approx 1/\lambda$ of opposite signs, that
   cancel in the sum. }
    \vspace{.5 cm}
   \begin{tabular}{|c|c|c|c|c|c|}
   \hline
   $\lambda $ &  $E/m$    &$\Re F_+(IDR)$&$\Re F_+(sDDR)$&$\Delta^{+}(E,m)$&$\Re F_+$(full DDR) \\
     \hline
 0.1  & 2 & 0.403965347917&64.104711735696&-63.700746387780 & 0.403965347917\\
  & 10 &-1.157501083855&62.505923539517&-63.663424623364& -1.157501083847\\
  & 100 &-8.012794316442&55.649197356317&-63.661991672959&-8.012794316643\\
  & 1000 & -22.610812691731&41.051164689357&-63.661977381116&-22.610812691759\\
\hline
 0.01  & 2 & 0.335089941935&6366.575003517582&-6366.239913575625&0.335089941935\\
       & 10 &-1.167198982181&6365.032104422951&-6366.199303405104&-1.167198982174\\
       & 100&-6.387924518333&6359.809814915427&-6366.197739433735&-6.387924518332\\
       & 1000 &-15.306917952512&6350.890805880863&-6366.197723833389&-15.306917952549\\
\hline
 0.001  & 2 & 0.328764505695&636620.1436892521&-636619.8149247168&0.328764505695\\
        & 10 &-1.165818944785&636618.6081426198&-636619.7739615351&-1.165818944778\\
        & 100&-6.242924888218&636613.5294586228&-636619.7723834814&-6.242924888217\\
        & 1000 &-14.728077626309&636605.0442901437&-636619.7723677403&-14.728077626321\\
 \hline
  0     & 2     &0.328067590880&  ${\bf -\infty}$&${\bf +\infty}$&0.328067590880\\
        & 10    &-1.165643354601& ${\bf -\infty}$&${\bf +\infty}$ & -1.165643354595\\
        & 100   &-6.227003476713& ${\bf -\infty}$&${\bf +\infty}$ & -6.227003476712\\
        & 1000  &-14.665220640485& ${\bf -\infty}$&${\bf +\infty}$& -14.665220640497\\
 \hline
 \hline
    \end{tabular}
   \end{table}
   \end{center}

\begin{center}
   \begin{table}
\caption{ \label{tab4} Real amplitudes from $F_+$ of  form
${\rm Im}\, F_+= (E/m)^\lambda \ln^2(E/m) $  
  at several energies, with
    $\lambda$ approaching the critical value  $\lambda=0$. The standard 
    DDR contribution and  correction terms $\Delta^+$ added together 
    reproduce perfectly the direct calculation  of the  IDR principal value 
    integral (calculated using the Mathematica 6  software program). For 
    $\lambda \rightarrow 0$, sDDR and the corrections $\Delta^+$ have
   singularities of form $\approx 1/\lambda$ of opposite signs, that
   cancel in the sum. }
     \vspace{.5 cm}
   \begin{tabular}{|c|c|c|c|c|c|}
   \hline
   $\lambda $ &  $E/m$    &$\Re F_+(IDR)$&$\Re F_+(sDDR)$&$\Delta^{+}(E,m)$&$\Re F_+$(full DDR)\\
     \hline
 0.1  & 2  &0.825818895910&-1272.449349885946&1273.275168781857&0.825818895910  \\
      & 10 &0.385649697828&-1272.855271730595&1273.240921428424&0.385649697828\\
  & 100 & -19.862174910070&-1293.101733393810&1273.239558483750&-19.862174910061\\
  & 1000 &-98.840346530164&-1372.079891403041&1273.239544872647&-98.840346530396\\
\hline
 0.01  & 2 &0.708180406300&-1273238.877089482&1273239.585269881&0.708180406300\\
       & 10 &-0.132971707507&-1273239.679276770&1273239.546305056&-0.132971707508\\
       & 100 &-16.282484687019&-1273255.827235536&1273239.544750842&-16.282484687013\\
       & 1000&-65.648229253245&-1273305.192964580&1273239.544735319&-65.648229253369\\
\hline
 0.001  & 2 & 0.697502589569&-1.273239544078824E+09&1.273239544776237E+09&0.697502589569\\
        & 10 &-0.173398449022&-1.273239544910241E+09&1.273239544736754E+09&-0.173398449022\\
        & 100&-15.940304163616&-1.273239560675572E+09&1.273239544735179E+09&-15.940304163611\\
        & 1000&-63.000875389386&-1.273239607736127E+09&1.273239544735163E+09&-63.000875389502\\
     \hline
  0     & 2     &0.696327412793& ${\bf -\infty}$&${\bf +\infty}$ &0.696327412793\\
        & 10    &-0.177778228941& ${\bf -\infty}$&${\bf +\infty}$  & -0.177778228942\\
        & 100   &-15.902527098975& ${\bf -\infty}$&${\bf +\infty}$ & -15.902527098969\\
        & 1000  &-62.713313030736& ${\bf -\infty}$&${\bf +\infty}$ & -62.713313030851\\
 \hline
 \hline
    \end{tabular}
   \end{table}
   \end{center}

\end{document}